%% file: main_arxiv.tex
\documentclass{article}

\PassOptionsToPackage{table}{xcolor}
\usepackage{iclr2027_conference,times}
\input{math_commands.tex}

\usepackage{booktabs}
\usepackage{array}
\usepackage{xcolor}
\usepackage{tabularx}
\usepackage{graphicx}
\usepackage{amsmath}
\usepackage{amssymb}
\usepackage{enumitem}
\usepackage{float}
\usepackage{hyperref}
\usepackage{url}
\usepackage{iftex}
\ifXeTeX
  \usepackage[T1]{fontenc}
\fi
\hypersetup{hidelinks}

\iclrfinalcopy

\title{When Does a Spoken Agent Have Enough\\
Evidence to Act?\\
The PACT-SLM Contract Test}
\author{\textbf{Mengzhe Geng}\\
\normalfont National Research Council Canada\\
\normalfont \texttt{Mengzhe.Geng@nrc-cnrc.gc.ca}}

\renewcommand{\headrulewidth}{0pt}
\renewcommand{\footrulewidth}{0pt}
\fancypagestyle{plain}{%
  \fancyhf{}
  \fancyfoot[C]{\thepage}
  \renewcommand{\headrulewidth}{0pt}
  \renewcommand{\footrulewidth}{0pt}}
\begin{document}
\maketitle
\fancyhead{}
\fancyfoot[C]{\thepage}
\renewcommand{\headrulewidth}{0pt}
\renewcommand{\footrulewidth}{0pt}
\pagestyle{fancy}
\thispagestyle{plain}

\input{paper_body}

\begingroup
\small
\bibliographystyle{iclr2027_conference}
\bibliography{refs}
\endgroup

\appendix
\input{appendix}

\end{document}

%% file: math_commands.tex
\usepackage{amsmath,amsfonts,bm}

\def\eqref#1{equation~\ref{#1}}

\def\1{\bm{1}}

\DeclareMathAlphabet{\mathsfit}{\encodingdefault}{\sfdefault}{m}{sl}
\SetMathAlphabet{\mathsfit}{bold}{\encodingdefault}{\sfdefault}{bx}{n}



%% file: paper_body.tex
\begin{abstract}
Streaming spoken agents may produce the correct final action after acting too early. Final-turn scores do not reveal whether each observed speech prefix supports an exposed action. We introduce the Partial Speech Action Contract for Turn Taking in Speech Language Models (PACT-SLM), a controlled test that assigns the first valid action time and evaluates both action identity and timing. In the primary test, 80 paired contrast groups from four held-out semantic families yield 1,600 prefix predictions across clean and 15 dB noise renderings. Using source-utterance semantic targets rather than counterbalanced branch codes, WavLM Base Plus reaches 26.03\% pooled post-onset semantic-label accuracy (95\% group-bootstrap interval [22.14\%, 29.68\%]), exposes an action on 18.99\% of pre-onset prefixes, and predicts 5.94\% of complete trajectories exactly. Its pooled label score is at the 96th percentile of 100 within-prefix label permutations, below the 97.5th-percentile reference (26.73\%). It exceeds matched text, scalar-acoustic, and shuffled-representation probes in pooled post-onset label accuracy. Elapsed time is more onset-exact (36.25\% versus 23.13\%) but less accurate about action identity (9.92\% versus 26.03\%). These results motivate separate measurement of action identity and onset timing in partial-speech evaluations.
\end{abstract}

\section{Introduction}

Spoken agents must often decide before a user finishes speaking. They may continue listening, reason internally, ask for clarification, call a tool, or interrupt. A final-response score can verify the last action and still miss the central online error: the system acted before the speech available at that moment supported the action. That distinction matters whenever an exposed action changes external state or interrupts the user.

Current speech language models (SLMs) and full-duplex systems make early action possible, but model capability does not define a valid timing test \citep{chiang2025shanks,chiang2025stitch,defossez2024moshi,ge2025flexi,lin2025fullduplexbenchv2}. The evaluator must know which prefix was visible, when an action first became valid, and whether a model used action-bearing evidence or a shortcut such as elapsed duration. Without this contract, high final accuracy can reward a system that guesses the action late, acts early, or simply recognizes an endpoint.

We ask: \emph{what must an evaluation control to determine whether partial speech contains enough evidence for action?} PACT-SLM answers with a paired construction\footnote{Prediction-level replay code and saved predictions: \url{https://github.com/MENGZHEGENG/pact-speech-action-calibration}.}. Two generated utterances share a transcript and observed audio through the last prefix labeled \textsc{Wait}, then diverge before the first prefix at which their different actions become valid. All later prefixes are nested crops of one branch waveform. Training uses clean speech; testing holds out both semantic families and a 15 dB noise rendering. Matched controls expose duration, scalar-acoustic, text, and representation shortcuts.

Branch IDs do not encode semantic actions, so targets come from the source utterances. WavLM Base Plus exceeds matched controls in post-onset accuracy but falls below the 97.5th-percentile permutation reference and exposes actions before onset. Elapsed time identifies onset more often but action identity less often, motivating separate measures.

\paragraph{Contributions.}
\begin{enumerate}[leftmargin=*,itemsep=1pt,topsep=2pt]
  \item We define a variable-onset evaluation contract that keeps paired pre-onset observations identical and preserves nested prefixes when noise is added.
  \item We instantiate the contract as a 340-group generated-speech diagnostic with five prefixes per trajectory and matched controls for endpoint, elapsed time, authored text, scalar acoustics, and shuffled speech representations.
  \item We evaluate 80 held-out groups using the semantic action associated with each source utterance, rather than its counterbalanced branch code. WavLM Base Plus exceeds matched controls in pooled post-onset label accuracy but falls below the 97.5th-percentile permutation reference, exposes actions before onset, and predicts few complete trajectories. These results motivate separate measurement of action identity and timing.
\end{enumerate}

\section{Related work and evaluation gap}

Speech encoders and modality adapters provide input interfaces for SLMs, including WavLLM, COSMIC, and AlignFormer \citep{hu2024wavllm,pan2024cosmic,fan2025alignformer}. Simultaneous and full-duplex models address listening, reasoning, speaking, and interruption during an ongoing turn \citep{chiang2025shanks,chiang2025stitch,defossez2024moshi}. Broader duplex benchmarks evaluate conversation quality and turn management \citep{ge2025flexi,lin2025fullduplexbenchv2}. PACT-SLM complements these systems by testing whether the current user-speech prefix supports an exposed action.

Prosody and contextual speech benchmarks evaluate paralinguistic or semantic reasoning \citep{qian2025prosodylm,wang2025cpbench,wang2026emotionthinker}. Their examples can reveal whether a representation encodes acoustic information, but an action-timing evaluation also needs a known onset, strictly nested observations, and controls that separate action evidence from time. We use frozen WavLM representations \citep{chen2022wavlm} as a reproducible probe, not as a new SLM architecture.

\section{PACT-SLM evaluation contract}
\label{sec:contract}

\subsection{What the test must identify}

Let trajectory $i$ have ordered speech prefixes $p_{i,1},\ldots,p_{i,K}$, target action $a_i$, and first valid action index $k_i^*$. The target at prefix $k$ is
\begin{equation}
y_{i,k} =
\begin{cases}
\textsc{Wait}, & k < k_i^*,\\
a_i, & k \geq k_i^*.
\end{cases}
\label{eq:trajectory-target}
\end{equation}
The evaluation must therefore answer two questions. Does the model identify $a_i$ after the relevant evidence appears? Does it keep waiting before $k_i^*$ and change state at the correct prefix? Final-action accuracy answers only the first question at $k=K$.

\subsection{Paired branch construction}

\begin{figure}[H]
  \centering
  \includegraphics[width=\textwidth]{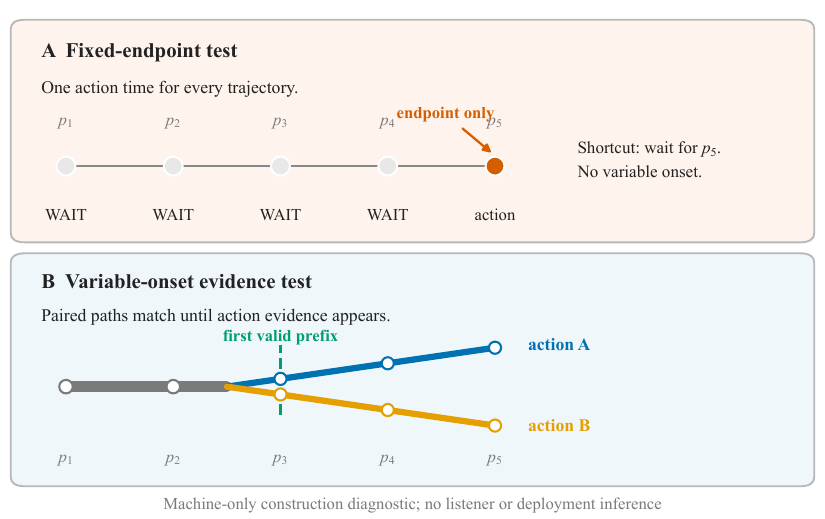}
  \caption{The PACT-SLM contract test. In the fixed-endpoint test (top), an action is valid only at the final prefix, so a duration or endpoint rule can wait until the last crop. In the variable-onset test (bottom), paired observed prefixes remain identical through the last required wait, action-specific evidence appears before the first valid action prefix, and every clean or noisy prefix is cropped from one branch waveform. Held-out families and noise test whether a probe uses transferable action evidence.}
  \label{fig:variable-onset-contract}
\end{figure}

Figure~\ref{fig:variable-onset-contract} contrasts a fixed-endpoint shortcut with the PACT-SLM test. Each contrast group contains two transcript-matched source utterances with different actions. We assign $k_i^*$ to prefix fraction 0.4, 0.6, or 0.8, balanced within split and action-pair buckets. The source waveforms begin to diverge at the temporal midpoint between the endpoint of the preceding observed prefix and the endpoint of the first action-labeled prefix. Before that frame, the two branch waveforms use the same averaged samples. A 50 ms crossfade then connects each branch to its own source waveform. The observed prefixes before onset are byte-identical within a pair, while the first action-labeled prefix contains branch-specific samples. This first-valid index is assigned by the construction; it is not a listener judgment of when an action becomes appropriate.

Each branch yields five nested crops at fractions $\{0.2,0.4,0.6,0.8,1.0\}$. In the noisy condition, one pair-level Gaussian-noise realization (15 dB signal-to-noise ratio) is added to both utterances before cropping. The construction tests a controlled information contract, not perceived turn naturalness.

\subsection{Data and split}

The generated-speech corpus contains transcript-matched action contrasts. We excluded 120 pairs because one or both referenced source waveforms were unavailable. Of the 360 pairs with audio, 340 have two distinct target actions and enter the diagnostic; the other 20 same-action pairs do not form an action contrast. Here a \emph{contrast group} is one eligible pair of source utterances, a \emph{source trajectory} is one utterance in that pair, and a \emph{prefix record} is one of five crops of a source trajectory in one rendering. Thus, 208 training groups yield 2,080 clean prefix records; 52 development groups yield 520 clean records; and 80 test groups yield 1,600 records because both branches have five clean and five noisy crops. The total is 4,200 records from 680 distinct source trajectories, not 4,200 independent examples. Table~\ref{tab:canary-split} reports these units separately. The 80 contrast groups, each retaining both branches and renderings during resampling, are the bootstrap units.

The test split holds out four semantic families: book uncertainty, calendar noise, message barge-in, and timer urgency. Their 80 contrast groups do not occur in training or development. Clean speech is the only training and development rendering; test includes clean speech and the held-out noise rendering. Table~\ref{tab:canary-split} reports 13 family memberships for each of the training and development splits; overlap between those membership sets is unknown. The four test families are held out from both splits, and the clean test slice provides a matched reference for the noisy rendering.

\input{results_tables}

\section{Models and measurements}
\label{sec:models}

\paragraph{Frozen speech representation.}
We extract the final hidden state from the TorchAudio \texttt{WAVLM\_BASE\_PLUS} checkpoint for every prefix and mean-pool it over time. The resulting 768-dimensional vector is standardized using training data. The checkpoint has 94.4 million parameters and operates on 16 kHz audio; extraction used PyTorch 2.6.0, TorchAudio 2.6.0, and CUDA 12.4. The audio-only probe is balanced multinomial logistic regression with $C=1$, the \texttt{lbfgs} solver, random state 0, and a 2,000-iteration limit. No model or threshold is selected on the test set. The audio-plus-text probe concatenates the standardized speech vector with term frequency--inverse document frequency (TF--IDF) word and bigram features.

\paragraph{Matched controls.}
Always-wait measures the cost of never acting. Endpoint plurality receives an oracle final-prefix indicator and emits the most frequent training action only at that endpoint. Elapsed-only uses prefix duration. Text-only uses an authored token-fraction transcript, which is a controlled text feature and not a time-aligned transcript. Scalar-audio uses duration, root-mean-square energy, mean absolute amplitude, zero-crossing rate, and crest factor. For each of the five prefix positions, prefix-shuffled WavLM permutes the correspondence between training speech embeddings and action labels across examples with seed 0. This preserves position-specific action frequencies while breaking the speech--action association. Every learned probe uses the same classifier and training rows.

\paragraph{Measurements and uncertainty.}
Trajectory exactness requires all five actions for one source and rendering to match Equation~\ref{eq:trajectory-target}. Pre-onset exposed-action rate counts \textsc{Ask-Clarify}, \textsc{Call-Tool}, or \textsc{Interrupt-Safe} before $k_i^*$; internal \textsc{Think} and \textsc{Abstain} are not counted as exposed actions. \textsc{Speak} is excluded from the six scored labels. Pooled post-onset semantic-label accuracy scores the source-utterance action at and after $k_i^*$ for exposed-action and internal-state prefixes.

\newpage
\paragraph{Onset and uncertainty.}
Onset exactness checks whether the first non-\textsc{Wait} prediction occurs at $k_i^*$. We report 1,000 group-bootstrap repeats over the 80 contrast groups and paired intervals for model differences.

\paragraph{Analysis scope.}
The reported WavLM estimates come from probes fit on training embeddings using the source-utterance action labels defined above. Group-bootstrap intervals condition on these fitted probes and their held-out predictions; they do not include variability from feature extraction or probe fitting. The generated speech and embeddings are not included in the accompanying materials.

\section{Results}
\label{sec:results}

\subsection{Action identity and onset are different capabilities}

Table~\ref{tab:variable-onset-results} gives the full-test comparison under the source-utterance targets. WavLM Base Plus reaches 26.03\% pooled post-onset semantic-label accuracy [22.14\%, 29.68\%], predicts 5.94\% of complete trajectories exactly [3.44\%, 9.06\%], and exposes an action on 18.99\% of prefixes strictly before the assigned onset [14.19\%, 24.85\%]. The pooled post-onset score is 252/968: 195/606 = 32.18\% on exposed-action labels and 57/362 = 15.75\% on internal-state labels. It therefore is not an exposed-action-only score. These intervals resample the 80 contrast groups while holding the fitted probes and their held-out predictions fixed; they do not include fitting uncertainty. The early-exposure denominator contains only prefixes before onset, not all 1,600 test records. The result is evidence of imperfect action and timing behavior in this probe, not a reliable timing policy.

The matched controls qualify this result. WavLM Base Plus exceeds prefix-shuffled WavLM by an absolute difference of 10.74\% in pooled post-onset label accuracy [6.73\%, 14.56\%], and scalar acoustics by 11.16\% [5.66\%, 15.95\%]. However, in 100 within-prefix permutations of the training-label correspondence, the observed 26.03\% score is at the 96th percentile and below the 97.5th-percentile value of 26.73\%. The probe also exceeds the text-only control by 7.44\% [2.62\%, 12.60\%], but this permutation check does not support a robust speech-specific effect at the stated reference. These results are consistent with a limited action signal in this fixed generated-speech test; they do not isolate a general speech-representation effect.

Elapsed time exposes a separate tradeoff. It reaches 36.25\% onset exactness, compared with 23.13\% for WavLM, a WavLM-minus-elapsed absolute difference of -13.12\% [-23.76\%, -1.88\%]. Elapsed-only post-onset accuracy is 9.92\%, compared with 26.03\% for WavLM, a difference of 16.12\% [10.85\%, 21.50\%]. Thus, duration predicts the assigned onset more often but identifies the target action less often. Neither score alone describes the joint timing-and-action behavior.

\begin{figure}[H]
  \centering
  \includegraphics[width=\textwidth]{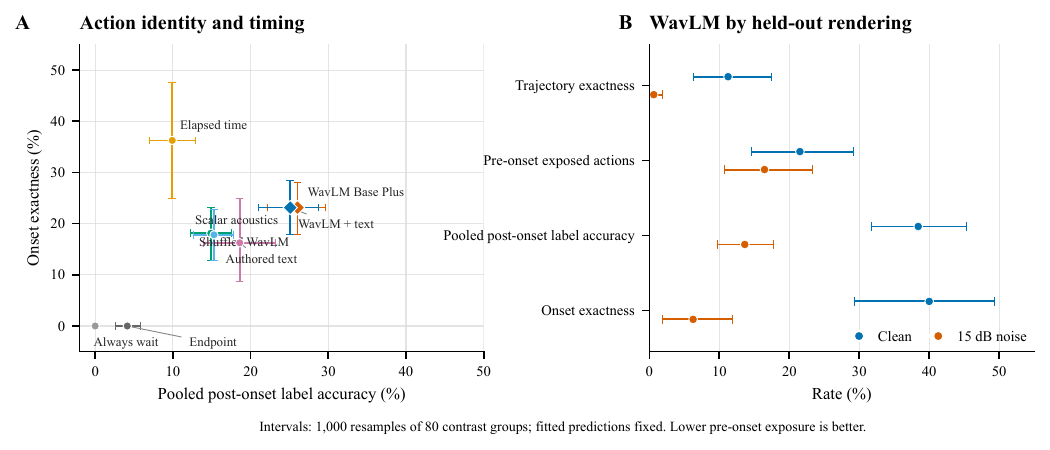}
  \caption{Results on the 80-group held-out test under source-utterance semantic targets. Each panel reports one metric for all scored probes; points and intervals are percentages and 95\% intervals from 1,000 resamples of the 80 contrast groups. Intervals hold fitted probes and their held-out predictions fixed. The audio-only probe exceeds matched learned controls on pooled post-onset label accuracy, while elapsed time has higher onset exactness. Complete-trajectory accuracy remains low.}
  \label{fig:identity-timing-results}
\end{figure}

Figure~\ref{fig:identity-timing-results} displays estimates and group-bootstrap intervals for all four outcomes. WavLM Base Plus has higher pooled post-onset label accuracy than the matched learned controls, while elapsed time has higher onset exactness. Under held-out noise, WavLM's pooled post-onset label accuracy falls from 38.43\% [31.76\%, 45.38\%] on clean speech to 13.64\% [9.80\%, 17.72\%]; complete-trajectory exactness falls from 11.25\% [6.25\%, 17.50\%] to 0.62\% [0.00\%, 1.88\%]. These clean/noise intervals resample groups, and the same groups contribute both renderings.

\subsection{Held-out noise breaks complete trajectories}

Table~\ref{tab:renderer-results} shows the rendering split. WavLM trajectory exactness is 11.25\% on clean speech and 0.62\% under held-out 15 dB noise. Pooled post-onset label accuracy falls from 38.43\% to 13.64\%, and onset exactness falls from 40.00\% to 6.25\%. Adding authored text changes the combined pre-onset exposed-action rate from 18.99\% to 15.19\%, an absolute difference of -3.80\% for the text-augmented probe minus the audio-only probe [-6.00\%, -1.94\%]. Its pooled post-onset label difference is -0.93\% [-2.06\%, 0.10\%] and its trajectory-exactness difference is 0.63\% [-0.31\%, 1.88\%]. In this comparison, text reduces early exposed actions but does not improve action identification; the tradeoff is not an overall gain.

\begin{table}[H]
\caption{WavLM Base Plus results by held-out rendering. Values are percentages with 1,000 contrast-group-bootstrap 95\% intervals.}
\label{tab:renderer-results}
\centering
\small
\setlength{\tabcolsep}{4pt}
\begin{tabularx}{\columnwidth}{>{\raggedright\arraybackslash}Xrr}
\toprule
\textbf{Metric} & \textbf{Clean} & \textbf{15 dB noise} \\
\midrule
Trajectory exact $\uparrow$ & 11.25 [6.25, 17.50] & 0.62 [0.00, 1.88] \\
Pre-onset exposed $\downarrow$ & 21.52 [14.57, 29.17] & 16.46 [10.69, 23.32] \\
Pooled post-onset label $\uparrow$ & 38.43 [31.76, 45.38] & 13.64 [9.80, 17.72] \\
Onset exact $\uparrow$ & 40.00 [29.38, 49.38] & 6.25 [1.88, 11.88] \\
\bottomrule
\end{tabularx}
\end{table}

\subsection{Why the contract changes the conclusion}

Appendix~\ref{sec:prior-design-diagnostics} describes two endpoint-only diagnostics. A single-chunk design has no \textsc{Wait} target, and repeated templates allow a condition lookup to reach 93.75\% on development and test. A fixed-endpoint design adds prefixes but marks every action as valid only at the final crop, allowing an endpoint rule to avoid all premature-action errors. Neither design measures whether an action is supported at an intermediate prefix.

The variable-onset diagnostic evaluates both action identity and timing in one trajectory. Paired prefixes are identical before onset, and each post-onset target is the source utterance's semantic action. WavLM has low complete-trajectory accuracy and falls below the 97.5th-percentile permutation reference. These results do not show that speech contains no action information; they show that this probe and test do not support a robust speech-specific conclusion or a reliable joint timing policy.

\section{Scope and limitations}

PACT-SLM is a controlled diagnostic construction, not a natural-speech benchmark or deployed policy evaluation. Its waveforms join an averaged shared prefix to one of two generated branches, and no listener study establishes naturalness, intelligibility, or human agreement with the action onset. Six training-development families were excluded because 120 complete source pairs lacked audio; another 20 audio-backed pairs had no action contrast. The four test families were used during development, so their held-out results are exploratory rather than independent confirmation on untouched families.

The evaluated model uses a frozen WavLM Base Plus representation with a shallow classifier. The authored text control is truncated by token fraction and is not a timestamped automatic transcript. The results concern this representation and construction. Native streaming SLMs, timestamped transcripts, natural speech, additional voices and synthesis systems, and human action-appropriateness judgments require separate evidence.

\section{Conclusion}

PACT-SLM makes an evaluation question explicit: recognizing an eventual action and timing its exposure are distinct measurements. In the controlled generated-speech test, WavLM Base Plus reaches 26.03\% pooled post-onset semantic-label accuracy, but its score is below the 97.5th-percentile reference from 100 within-prefix permutations. It exposes actions before onset and predicts only 5.94\% of complete trajectories exactly; elapsed time is more onset-exact. The evidence supports using separate action and timing measures in this controlled generated-speech test. It does not establish a robust speech-specific effect, a reliable policy, or performance on natural conversation.

%% file: results_tables.tex
\begin{table*}[t]
\caption{Split composition after excluding pairs with unavailable source audio and pairs without an action contrast. A group is a paired action contrast; a source trajectory is one utterance in that group; a prefix record is one scored crop in one rendering. The 13 family memberships listed for training and development may overlap. Test includes clean and held-out 15 dB noise renderings.}
\label{tab:canary-split}
\centering
\small
\setlength{\tabcolsep}{5pt}
\begin{tabularx}{0.96\textwidth}{>{\raggedright\arraybackslash}Xrrrr>{\raggedright\arraybackslash}p{0.9in}}
\toprule
\textbf{Split} & \shortstack{\textbf{Family}\\\textbf{memberships}} & \textbf{Groups} & \shortstack{\textbf{Source}\\\textbf{trajectories}} & \shortstack{\textbf{Prefix}\\\textbf{records}} & \textbf{Rendering} \\
\midrule
Train & 13 & 208 & 416 & 2,080 & Clean \\
Development & 13 & 52 & 104 & 520 & Clean \\
Test & 4 & 80 & 160 & 1,600 & Clean, 15 dB noise \\
\bottomrule
\end{tabularx}
\end{table*}

\begin{table*}[t]
\caption{Variable-onset results on 80 held-out contrast groups across clean and 15 dB noise renderings (1,600 prefix predictions). Values are percentages. Trajectory exactness requires all five decisions for one trajectory and rendering to match. Exposed-action rate is calculated over pre-onset prefixes only and is lower when better; pooled post-onset semantic-label accuracy and onset exactness are higher when better. WavLM Base Plus uses source-utterance semantic targets. Intervals are in Appendix~\ref{sec:full-intervals}.}
\label{tab:variable-onset-results}
\centering
\small
\setlength{\tabcolsep}{4pt}
\begin{tabularx}{\textwidth}{>{\raggedright\arraybackslash}Xrrrr}
\toprule
\textbf{Model or control} & \shortstack{\textbf{Trajectory exact}\\\textbf{(\%) $\uparrow$}} & \shortstack{\textbf{Pre-onset exposed}\\\textbf{(\%) $\downarrow$}} & \shortstack{\textbf{Post-onset label}\\\textbf{(\%) $\uparrow$}} & \shortstack{\textbf{Onset exact}\\\textbf{(\%) $\uparrow$}} \\
\midrule
Always wait & 0.00 & 0.00 & 0.00 & 0.00 \\
Endpoint plurality & 0.00 & 0.00 & 4.13 & 0.00 \\
Elapsed time only & 0.00 & 4.43 & 9.92 & 36.25 \\
Authored text only & 3.75 & 58.23 & 18.60 & 16.25 \\
Scalar acoustics & 1.56 & 10.13 & 14.88 & 18.13 \\
Prefix-shuffled WavLM & 0.31 & 51.58 & 15.29 & 17.81 \\
\rowcolor{blue!8} WavLM Base Plus & 5.94 & 18.99 & 26.03 & 23.13 \\
\rowcolor{blue!8} WavLM Base Plus + authored text & 6.56 & 15.19 & 25.10 & 23.13 \\
\bottomrule
\end{tabularx}
\end{table*}

%% file: appendix.tex
\section{Sample units and split}
This appendix defines the sample units and uncertainty calculation for the variable-onset generated-speech diagnostic. The endpoint-only diagnostics below are separate designs and do not contribute observations to the primary comparison.

\paragraph{Counts and units.}
One contrast group contains two source utterances (two source trajectories); each trajectory has five nested prefix crops; each rendering is applied to the complete trajectory before cropping; and each prefix record is one crop under one rendering. The variable-onset diagnostic has 208 training groups (416 trajectories, 2,080 clean prefix records), 52 development groups (104 trajectories, 520 clean records), and 80 test groups (160 trajectories, 1,600 clean and 15 dB noise records). Thus, 4,200 records are repeated observations from 340 groups, not 4,200 independent examples. The group is the unit resampled for the reported intervals. Training and development each list 13 family memberships; their overlap is unknown. Four test families are held out from both.

\section{Limits of endpoint-only diagnostics}
\label{sec:prior-design-diagnostics}
The single-chunk design has no \textsc{Wait} target, while the fixed-endpoint design marks an action as valid only at the final crop. These designs show why final-action accuracy cannot determine whether the current prefix supports an action. Their observations are separate from the 340-group variable-onset study.

The single-chunk design scores one 1.5-second crop per utterance and has no \textsc{Wait} target, so it cannot measure action timing. Its splits contain repeated templates: 42 train--development, 126 train--test, and 42 development--test overlaps. A condition lookup reaches 93.75\% on both development and test, and adding intent raises the score to 100\%. These metadata lookups do not measure acoustic action recognition.

The fixed-endpoint design contains 64 contrast groups: 40 training, 8 development, and 16 test groups. Each group contributes two source trajectories, five crops per trajectory, and two renderings, for 1,280 prefix records. Every action becomes valid only at the final crop, so an endpoint rule is told when it may act. Noise is added independently after cropping, and the authored transcript is truncated by token fraction. Endpoint plurality reaches 37.50\% final-action accuracy with no pre-final non-\textsc{Wait} prediction. WavLM reaches 39.06\% final-action accuracy but predicts a non-\textsc{Wait} label on 25.78\% of pre-final records. Because the design has no intermediate valid onset, these results cannot measure whether an action is justified at a partial-speech prefix.

\section{Semantic action targets}
\label{sec:semantic-targets}
The branch codes \texttt{BRANCH\_A} and \texttt{BRANCH\_B} are counterbalanced across contrast groups and do not name fixed semantic actions. Each source trajectory therefore takes its action target from the corresponding source utterance. A branch identifier alone cannot serve as an action target because a probe could learn branch convention instead of semantic action.

Equation~\ref{eq:trajectory-target} assigns \textsc{Wait} to every prefix strictly before the valid onset and the source utterance's semantic action at and after onset. We fit the reported probes on training embeddings under these targets and evaluate their held-out predictions. The resulting estimates describe this embedding analysis; they do not include variability from new speech generation, feature extraction, or probe fitting.

\section{Full held-out results and uncertainty}
\label{sec:full-intervals}

Table~\ref{tab:full-model-intervals} reports the estimate and 95\% group-bootstrap interval for each model and each outcome. The inference unit is the contrast group: each resampled draw retains the two paired source trajectories, all available prefixes, and both test renderings. There are 80 groups and 1,000 bootstrap draws, with random seed 0. The fitted probes and their test predictions remain fixed during resampling. These intervals quantify variation across the observed test groups conditional on those predictions; they do not include uncertainty from fitting, new generated speech, a new synthesis system, or natural conversations.

The important pattern is not a uniformly strong audio result. WavLM Base Plus has 26.03\% post-onset accuracy [22.14\%, 29.68\%], but exactness across all five decisions of a trajectory is 5.94\% [3.44\%, 9.06\%]. Before onset, it exposes an action on 18.99\% of eligible prefixes [14.19\%, 24.85\%]. Elapsed time is less accurate about the action (9.92\% [6.96\%, 12.86\%]) but more often identifies the assigned onset (36.25\% [25.00\%, 47.50\%]). Hence, the best action-identity score and best onset score belong to different controls, and neither approximates a reliable joint policy.

\begin{table}[H]
\caption{Estimates in percent with 1,000 group-bootstrap 95\% intervals. Each cell is estimate [lower, upper]. The pre-onset exposure denominator contains only prefixes strictly before the valid onset.}
\label{tab:full-model-intervals}
\centering
\footnotesize
\setlength{\tabcolsep}{3pt}
\begin{tabularx}{\textwidth}{>{\raggedright\arraybackslash}p{0.95in}*{4}{>{\centering\arraybackslash}X}}
\toprule
\textbf{Model or control} & \textbf{Trajectory exact $\uparrow$} & \textbf{Pre-onset exposed $\downarrow$} & \textbf{Post-onset label $\uparrow$} & \textbf{Onset exact $\uparrow$} \\
\midrule
Always wait & 0.00 [0.00, 0.00] & 0.00 [0.00, 0.00] & 0.00 [0.00, 0.00] & 0.00 [0.00, 0.00] \\
Endpoint plurality & 0.00 [0.00, 0.00] & 0.00 [0.00, 0.00] & 4.13 [2.63, 5.86] & 0.00 [0.00, 0.00] \\
Elapsed time only & 0.00 [0.00, 0.00] & 4.43 [1.79, 7.69] & 9.92 [6.96, 12.86] & 36.25 [25.00, 47.50] \\
Text only & 3.75 [1.25, 6.88] & 58.23 [48.66, 68.02] & 18.60 [13.93, 23.14] & 16.25 [8.75, 25.00] \\
Scalar acoustics & 1.56 [0.31, 3.12] & 10.13 [7.10, 13.81] & 14.88 [12.24, 17.60] & 18.13 [12.81, 23.13] \\
Prefix-shuffled WavLM & 0.31 [0.00, 0.94] & 51.58 [46.73, 56.44] & 15.29 [12.70, 17.83] & 17.81 [12.81, 22.81] \\
\rowcolor{blue!8} WavLM Base Plus & 5.94 [3.44, 9.06] & 18.99 [14.19, 24.85] & 26.03 [22.14, 29.68] & 23.13 [17.81, 28.13] \\
\rowcolor{blue!8} WavLM Base Plus + authored text & 6.56 [3.75, 10.00] & 15.19 [10.60, 20.68] & 25.10 [21.07, 28.79] & 23.13 [17.81, 28.44] \\
\bottomrule
\end{tabularx}
\end{table}

\section{Paired differences and competing metrics}
\label{sec:paired-differences}

Table~\ref{tab:paired-differences} gives WavLM Base Plus minus each reference on the same 80 resampled groups. Differences are absolute differences in the named percentage-valued metric, computed from the paired group resamples; positive exposed-action differences mean more premature external actions and are unfavorable. Against prefix-shuffled WavLM, pooled post-onset label accuracy is higher by 10.74\% [6.73\%, 14.56\%]. Against scalar acoustics, it is higher by 11.16\% [5.66\%, 15.95\%]. These paired intervals quantify differences conditional on the fixed predictions; the separate permutation analysis below asks whether the score is unusual under altered training-label correspondence.

Elapsed time reverses the ranking for onset exactness: WavLM minus elapsed time is -13.12\% [-23.76\%, -1.88\%]. This contrast illustrates why it would be misleading to say that WavLM is simply better overall. Text augmentation also reverses some metrics. Relative to audio only, audio plus authored text lowers pre-onset exposure by 3.80\% [-6.00\%, -1.94\%], but its pooled post-onset label difference is -0.93\% [-2.06\%, 0.10\%] and its trajectory-exactness difference is 0.63\% [-0.31\%, 1.88\%]. Text changes the early-exposure outcome without a supported improvement in action identification.

\begin{table}[H]
\caption{Absolute differences in each named metric (\%) for WavLM Base Plus minus the named reference, with 95\% group-bootstrap intervals. Positive values favor WavLM for the upward metrics; a positive pre-onset exposure difference is unfavorable.}
\label{tab:paired-differences}
\centering
\footnotesize
\setlength{\tabcolsep}{3pt}
\begin{tabularx}{\textwidth}{>{\raggedright\arraybackslash}p{1.0in}*{4}{>{\centering\arraybackslash}X}}
\toprule
\textbf{Reference} & \shortstack{\textbf{$\Delta$ trajectory}\\\textbf{exact}} & \shortstack{\textbf{$\Delta$ pre-onset}\\\textbf{exposed}} & \shortstack{\textbf{$\Delta$ post-onset}\\\textbf{label}} & \shortstack{\textbf{$\Delta$ onset}\\\textbf{exact}} \\
\midrule
Always wait & +5.94 [+3.44, +9.06] & +18.99 [+14.19, +24.85] & +26.03 [+22.14, +29.68] & +23.13 [+17.81, +28.13] \\
Endpoint plurality & +5.94 [+3.44, +9.06] & +18.99 [+14.19, +24.85] & +21.90 [+19.02, +24.90] & +23.13 [+17.81, +28.13] \\
Elapsed time & +5.94 [+3.44, +9.06] & +14.56 [+8.75, +21.39] & +16.12 [+10.85, +21.50] & -13.12 [-23.76, -1.88] \\
Text only & +2.19 [-1.56, +5.94] & -39.24 [-49.68, -28.02] & +7.44 [+2.62, +12.60] & +6.88 [-2.82, +15.31] \\
Scalar acoustics & +4.38 [+1.25, +8.12] & +8.86 [+2.90, +15.50] & +11.16 [+5.66, +15.95] & +5.00 [-0.94, +11.56] \\
Shuffled WavLM & +5.62 [+2.81, +8.75] & -32.59 [-38.26, -26.77] & +10.74 [+6.73, +14.56] & +5.31 [-0.31, +10.94] \\
Audio + text & +0.63 [-0.31, +1.88] & -3.80 [-6.00, -1.94] & -0.93 [-2.06, +0.10] & 0.00 [-2.19, +2.19] \\
\bottomrule
\end{tabularx}
\end{table}

\paragraph{Permutation check.}
The audio-only post-onset score is 26.03\%. In 100 permutations of the training-label correspondence within prefix position (seeds 0--99), its empirical percentile is 96; the permutation median is 18.18\%, and the 97.5th-percentile reference is 26.73\%. Thus, the observed score does not exceed that reference. This finite permutation diagnostic uses one fixed split and one probe pipeline; it is not a calibrated population-level significance test and does not support a robust speech-specific claim. The permutation score is distinct from the 1,000 group-bootstrap intervals, which keep each model's predictions fixed.

\section{Results by held-out family}
\label{sec:family-results}

The pooled test score conceals variation across the four held-out families. Table~\ref{tab:family-results} reports descriptive point estimates for the audio-only and audio-plus-text probes. Each family contains 20 contrast groups, but the table's cell-wise confidence intervals are not a correction for inspecting multiple families. Pooled post-onset label accuracy is highest for message barge-in (47.08\% for audio only), while calendar noise is lowest (8.75\%). Exact full trajectories remain uncommon in every family: 0.00\% to 15.00\% for audio only. This variability means the pooled estimate should not be read as uniform transfer across semantic families.

\begin{table}[H]
\caption{Descriptive results within each of the four held-out semantic families (20 contrast groups per family). Values are percentages; these point estimates do not establish a between-family effect.}
\label{tab:family-results}
\centering
\footnotesize
\setlength{\tabcolsep}{1.5pt}
\begin{tabularx}{\textwidth}{>{\raggedright\arraybackslash}p{0.9in}*{6}{>{\centering\arraybackslash}X}}
\toprule
\textbf{Family} & \shortstack{\textbf{Audio}\\\textbf{post-onset}\\\textbf{label}} & \shortstack{\textbf{Text+audio}\\\textbf{post-onset}\\\textbf{label}} & \shortstack{\textbf{Audio exposed}\\\textbf{before}\\\textbf{onset}} & \shortstack{\textbf{Audio onset}\\\textbf{exact}} & \shortstack{\textbf{Audio}\\\textbf{trajectory}\\\textbf{exact}} & \shortstack{\textbf{Text+audio}\\\textbf{trajectory}\\\textbf{exact}} \\
\midrule
Book uncertainty & 19.67 & 17.62 & 14.10 & 21.25 & 0.00 & 1.25 \\
Calendar noise & 8.75 & 7.92 & 22.50 & 18.75 & 1.25 & 1.25 \\
Message barge-in & 47.08 & 47.50 & 12.50 & 22.50 & 15.00 & 16.25 \\
Timer urgency & 28.69 & 27.46 & 26.92 & 30.00 & 7.50 & 7.50 \\
\bottomrule
\end{tabularx}
\end{table}

\clearpage
\section{Construction and implementation details}
\label{sec:construction-details}

\paragraph{Waveform and prefix construction.}
Each contrast group pairs transcript-matched source utterances with different actions. The assigned valid-onset fraction is 0.4, 0.6, or 0.8, balanced within split and action-pair buckets. The source branches diverge at the temporal midpoint between the end of the preceding observed prefix and the end of the first action-labeled prefix. Before divergence, branch waveforms use the same averaged samples; a 50 ms crossfade connects each shared segment to its branch source. Prefixes are cropped at fractions $\{0.2,0.4,0.6,0.8,1.0\}$. In the noisy condition, one pair-level Gaussian-noise realization (15 dB signal-to-noise ratio) is added to both utterances before cropping, preserving nested prefixes. The construction tests an information contract, not perceived naturalness or listener agreement.

\paragraph{Probe and controls.}
The speech representation is the final hidden state from TorchAudio's \texttt{WAVLM\_BASE\_PLUS} checkpoint, mean-pooled over time into a 768-dimensional vector and standardized using training data. The classifier is balanced multinomial logistic regression with $C=1$, the \texttt{lbfgs} solver, random state 0, and a 2,000-iteration limit. Training and development use clean rendering only, and no threshold or model choice uses test scores. The audio-plus-text probe concatenates these standardized embeddings with word and bigram term frequency--inverse document frequency features. The text-only comparator uses authored token-fraction text rather than a time-aligned transcript. Scalar acoustics include duration, root-mean-square energy, mean absolute amplitude, zero-crossing rate, and crest factor. Prefix-shuffled WavLM permutes the pairing between training embeddings and action labels separately at each of the five prefix positions with seed 0; it retains position-specific action frequencies but removes the observed speech-to-action pairing. Endpoint plurality receives an oracle indicator of the final prefix and emits the most common training action only there. Always-wait never emits an action.

\paragraph{Metric definitions.}
For a trajectory, the target sequence is \textsc{Wait} at every prefix strictly before its assigned onset and the source-utterance semantic action at and after onset. Trajectory exactness is the fraction of source-trajectory/rendering combinations for which all five predictions match. Pre-onset exposed-action rate is the number of predictions equal to \textsc{Ask-Clarify}, \textsc{Call-Tool}, or \textsc{Interrupt-Safe} divided by the number of prefixes strictly before onset; internal \textsc{Think} and \textsc{Abstain} do not count as exposed actions. Pooled post-onset semantic-label accuracy is the number of correct source-utterance action labels divided by all prefixes at or after onset. For WavLM Base Plus, 252 of 968 post-onset predictions are correct. This includes 195/606 exposed-action labels (32.18\%) and 57/362 internal-state labels (15.75\%). Onset exactness is the fraction of source-trajectory/rendering combinations whose first non-\textsc{Wait} prediction occurs at the assigned onset; trajectories that never emit a non-wait action are assigned a predicted onset of 1.2 for timing-error calculations and do not count as exact. \textsc{Speak} is not a label in this six-class diagnostic.

\paragraph{Uncertainty and interpretation.}
For each of the 1,000 percentile-bootstrap draws, 80 contrast groups are sampled with replacement. All records belonging to a sampled group are retained together so the two paired utterances and both test renderings remain associated; duplicated draws are treated as separate bootstrap copies. Paired model differences reuse the same sampled group IDs for both models. The seed is 0 and the intervals are percentile quantiles at 2.5\% and 97.5\%. This procedure describes test-group resampling conditional on the fitted probes and their held-out predictions. It does not refit the probe in each draw. The 100-permutation distribution is a separate analysis and must not be conflated with bootstrap uncertainty.

\section{Interpretation boundaries}
\label{sec:interpretation-boundaries}

This is a machine-only probe diagnostic on synthesized speech with assigned onset labels. It does not measure human judgments of action appropriateness or speech naturalness, or test a native streaming speech-language model that acts in conversation. The observed score did not exceed the 97.5th percentile of the 100-permutation reference distribution. These results characterize this probe, not general properties of speech representations or agent timing.